\documentclass[aps,twocolumn,amsmath,amssymb,%
                pre,showpacs,array]{revtex4}

\usepackage{graphicx}
\usepackage{epsfig}

\begin{document}

\title{Separation of trajectories and its Relation to Entropy for Intermittent Systems with a Zero Lyapunov exponent}
\author{Nickolay Korabel and Eli Barkai}
\affiliation{Physics Department, Institute of Nanotechnology and Advanced Materials, Bar-Ilan University, Ramat-Gan 52900, Israel}

\date{\today}

\begin{abstract}

One dimensional intermittent maps with stretched exponential $\delta x_t \sim \delta x_0 e^{\lambda_{\alpha}t^{\alpha}}$ separation of nearby trajectories are considered. When $t \rightarrow \infty$ the standard Lyapunov exponent $\lambda= \sum_{i=0}^{t-1} \ln \left| M'(x_i) \right|/t$ is zero ($M'$ is a Jacobian of the map). We investigate the distribution of $\lambda_{\alpha}= \sum_{i=0}^{t-1} \ln \left| M'(x_i) \right|/t^{\alpha}$, where $\alpha$ is determined by the nonlinearity of the map in the vicinity of marginally unstable fixed points. The mean of $\lambda_{\alpha}$ is determined by the infinite invariant density. Using semi analytical arguments we calculate the infinite invariant density for the Pomeau-Manneville map, and with it obtain excellent agreement between numerical simulation and theory. We show that $\alpha \left< \lambda_{\alpha}\right>$ is equal to Krengel's entropy and to the complexity calculated by the Lempel-Ziv compression algorithm. This generalized Pesin's identity shows that $\left< \lambda_{\alpha}\right>$ and Krengel's entropy are the natural generalizations of usual Lyapunov exponent and entropy for these systems.

\end{abstract}


\maketitle

\section{Introduction}\label{Intro}

In recent years there is growing interest in dynamical systems which exhibit unpredictable behavior but are characterized by zero Lyapunov exponents, namely trajectories are separated non-exponentially \cite{Zasl,Kla,note}. Prominent examples of such systems are Hamiltonian models with a mixed phase space \cite{Zasl,ZE05}, systems with long range forces \cite{Latora}, certain billiards \cite{DC01,ACG97,ARV04,Li,CP99}, maps with discontinuities \cite{CNFV03}, one-dimensional hard-particle gas \cite{Gras}, maps at the edge of chaos (Feigenbaum's point) \cite{BR04,AT04}, generalized logistic map close to pitchfork and tangent bifurcation points \cite{BR02} and maps with marginal fixed points \cite{Grigo1,PM80,GW88,GT} such as Pomeau-Manneville map \cite{PM80} which was used to model intermittency. 

For the Pomeau-Manneville map the separation of trajectories is described by stretched exponentials $dx_t \sim dx_0 \; e^{\lambda_{\alpha}t^{\alpha}}$ with $0 < \alpha \le 1$ \cite{GW88}, which is related to power law sojourn times in the vicinity of marginally unstable fixed point (see details below). Systems with stretched exponential separation of trajectories behave very differently than normal chaotic ones so the standard description of dynamics is limited. Non-Gaussian fluctuations \cite{GW88}, aging \cite{Barkai03}, anomalous diffusion \cite{Zasl02,GT84,Zum93} and weak ergodicity breaking \cite{BB06,RB07,RB08,BB05} are found in these systems. Physical observables are described by distributions as for example given by Aaronson-Darling-Kac (ADK) and Dynkin-Lamperti theorems \cite{BB06,RB07,RB08,BB05,TZ06,Thaler02,Thaler98,Akimoto}. In particular Akimoto and Aizawa showed that the distribution of $\lambda_{\alpha}$ is given by the Mittag-Leffler function \cite{JKP}. Closely connected with the stretched exponential separation of trajectories, the Kolmogorov-Sinai (KS) entropy rate is zero since the entropy production increases sub-linearly in time \cite{GW88} and the relevant measure has infinite invariant density \cite{Aaronson,Thaler83,Thaler1} (see details below). Hence, there is a need to modify basic concepts of chaos theory when dealing with such systems. 


A fundamental theorem of chaos theory is Pesin's identity \cite{ER,Dor}. It states that for a closed chaotic system the KS-entropy is equal to the sum of positive Lyapunov exponents. For intermittent systems which we consider Lyapunov exponents are zero and so are their KS-entropies. Hence, the standard Pesin's identity is not useful. A natural question is whether Pesin's identity could be generalized by relating entropy concepts with the separation of trajectories for such systems. For the logistic map at the edge of chaos (Feigenbaum's point) with a power law separation of trajectories \cite{Gras81} a generalized Pesin's identity was investigated using Tsallis statistics \cite{BR04,AT04}. A critical discussion of this approach is given in Ref.\ \cite{Gra05} (and see a reply in \cite{Rob06}). Note that in this case the invariant density is a discontinuous fractal Cantor set \cite{ER} which is different from the absolutely continuous but infinite invariant densities inherent in our approach.

In this paper we are addressing the generalized Pesin's identity for intermittent maps with marginal fixed points. For the Pomeau-Manneville map an attempt in this direction, based on numerical simulations was made in \cite{Kor}. However, without the use of the infinite invariant density, and without the important concept of Krengel's entropy \cite{Krengel} (see details below) conclusions were limited. Recently we suggested a generalized Pesin's identity using the Pomeau-Manneville map \cite{KB09}. Particularly, it was shown that the average separation of trajectories is equal to Krengel's entropy $h_{\alpha}$: $h_{\alpha} = \alpha \left< \lambda_{\alpha} \right>$. The goal of this paper is to provide detailed derivation of generalized Pesin's identity, evidence for it for intermittent maps with one and two unstable fixed points, and importantly to clarify the meaning of entropy for systems with intermittency.

\section{Models}\label{models}

We consider two types of one-dimensional discrete time maps $x_{t+1} \equiv M(x_t)$ on the unit interval.

\begin{figure}[tbp]
\centerline{\psfig{figure=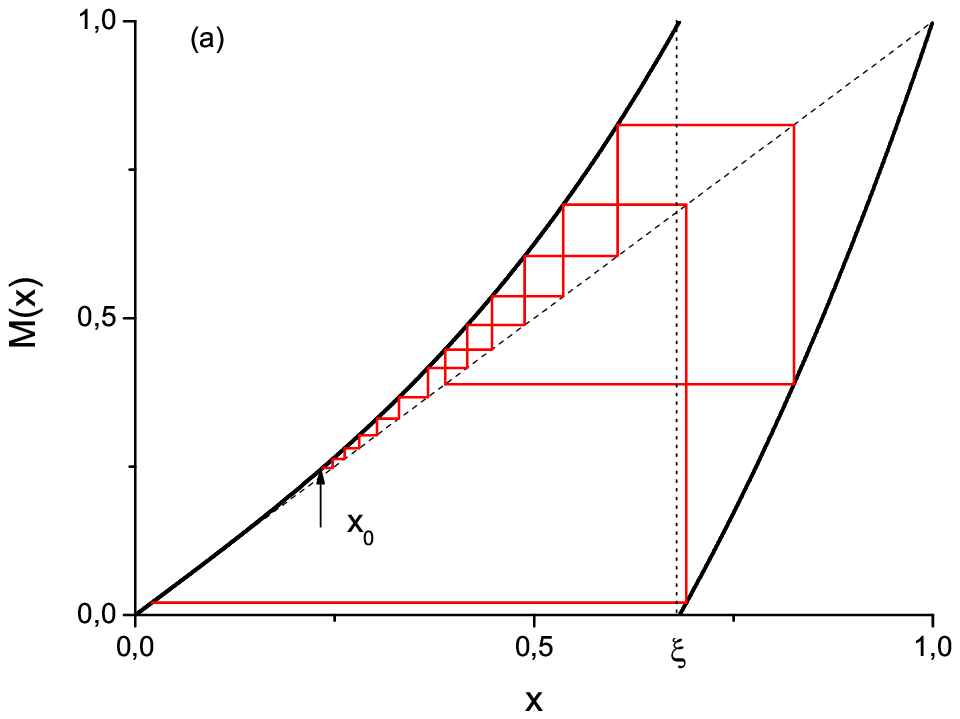,width=80mm,height=60mm}}
\centerline{\psfig{figure=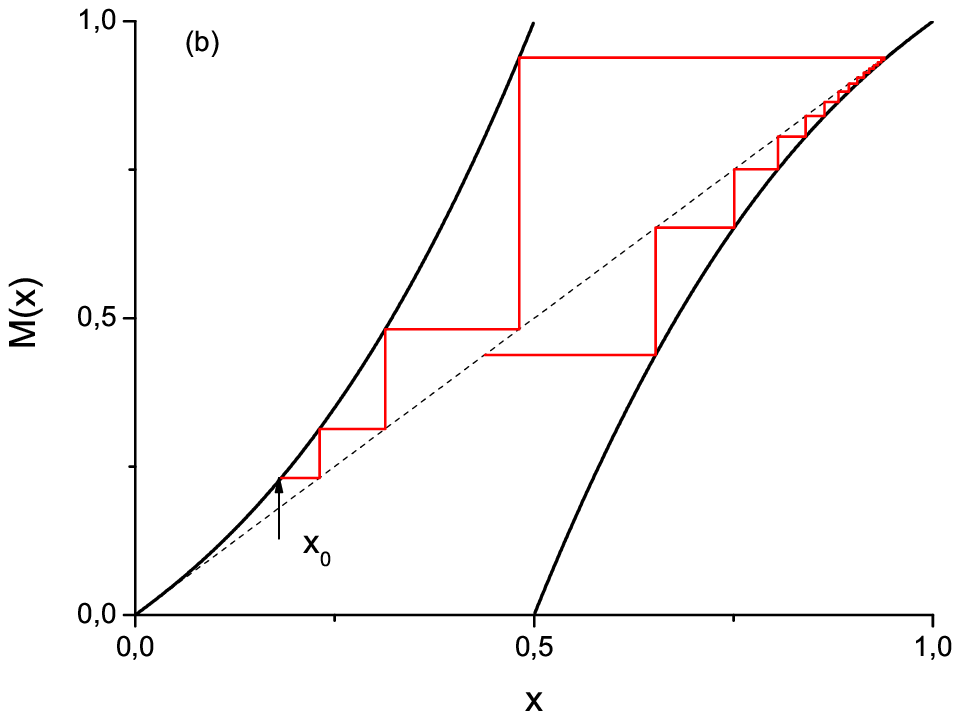,width=80mm,height=60mm}}
\caption{(color online) (a) Pomeau-Manneville map Eq.\ (\ref{map_eq_1}) with $z = 3$ and $a = 1$, so $\xi=0.6823$. Notice the unstable fixed point on $x = 0$. (b) The map in Eq.\ (\ref{map_eq_2}) with $z = 3$ has two unstable fixed points on $x = 0$ and $x = 1$. As usual $M'(x) = 1$ for marginally unstable fixed points.}
\label{Fig_Manneville_map}
\end{figure}

1) The first model is the Pomeau-Manneville map \cite{PM80}
\begin{equation}
\label{map_eq_1} 
M(x_t) =  x_t + a x_{t}^{z}  \; \left(\text{mod.} 1 \right), \; \; z\ge1, \; a>0.
\end{equation}
The discontinuity point $\xi$ is given by $M(\xi)=1$. A sketch of the map is shown in Fig.\ \ref{Fig_Manneville_map} (a).

2) The second map is defined as
\begin{equation}
\label{map_eq_2} 
M(x_t) = x_t + \begin{cases} 2^{1-z} \; x_t^z, & 0 < x_t < \frac{1}{2} \cr
- 2^{1-z} \; (1-x_t)^z, & \frac{1}{2} < x_t < 1.
\end{cases}
\end{equation}
This map has two marginally unstable fixed points $x_t = 0$ and $x_t = 1$ as shown in Fig.\ \ref{Fig_Manneville_map} (b). Such intermittent systems were analyzed with many different methods such as stochastic modeling \cite{HHS,Grigo1}, renormalization group theory \cite{PS}, continuous time random walks \cite{GT84,Zum93}, periodic orbit theory \cite{Artuso}, thermodynamic formalism \cite{PrelSlaw} and others \cite{GT,Maes}. For both models the behavior in vicinities of marginally unstable fixed points is controlled by  
\begin{equation}
\label{alf} 
\alpha = \begin{cases} 1, & z < 2 \cr
\frac{1}{z-1}, & z \ge 2.
\end{cases}
\end{equation}
At $z=2$ there exists a transition from ergodic behavior to behavior characterized by weak ergodicity breaking \cite{BB06,RB07,RB08,BB05}. 

Remark: although the expansion of the higher order iteration of the generalized logistic map with parameter values close to tangent bifurcations, is close in form to the Pomeau-Manneville with nonlinearity parameter $z=2$ \cite{BR02}, $x_{n+1} \sim x_n + x_n^2 + \epsilon$ with $\epsilon \ne 0$ ($\epsilon = 0$ exactly at tangent bifurcations points), this system is quite different from the Pomeau-Manneville map. It has faster than exponential separation of trajectories \cite{BR02}. The main difference between two systems is that the invariant density in the former case is a discontinuous fractal Cantor set, while it is absolutely continuous but infinite for the later system \cite{Aaronson,Thaler83,Thaler1}. We derive the infinite invariant density for the Pomeau-Manneville map explicitly in Sec\ \ref{PDF}.

\section{Separation of trajectories}\label{Sep}

\begin{figure}[tbp]
\centerline{\psfig{figure=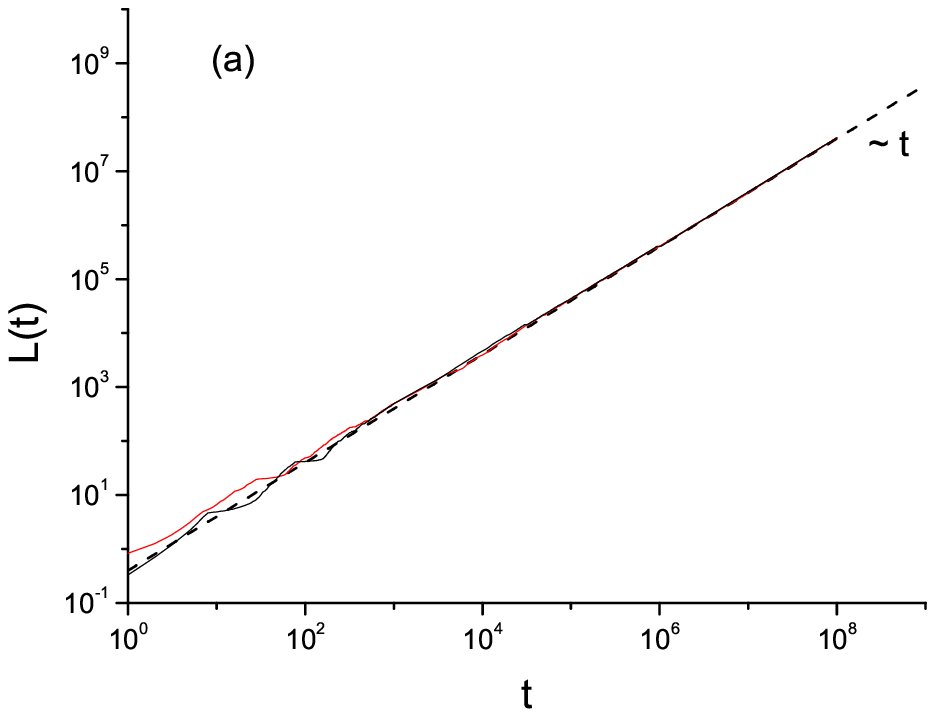,width=80mm,height=60mm}}
\centerline{\psfig{figure=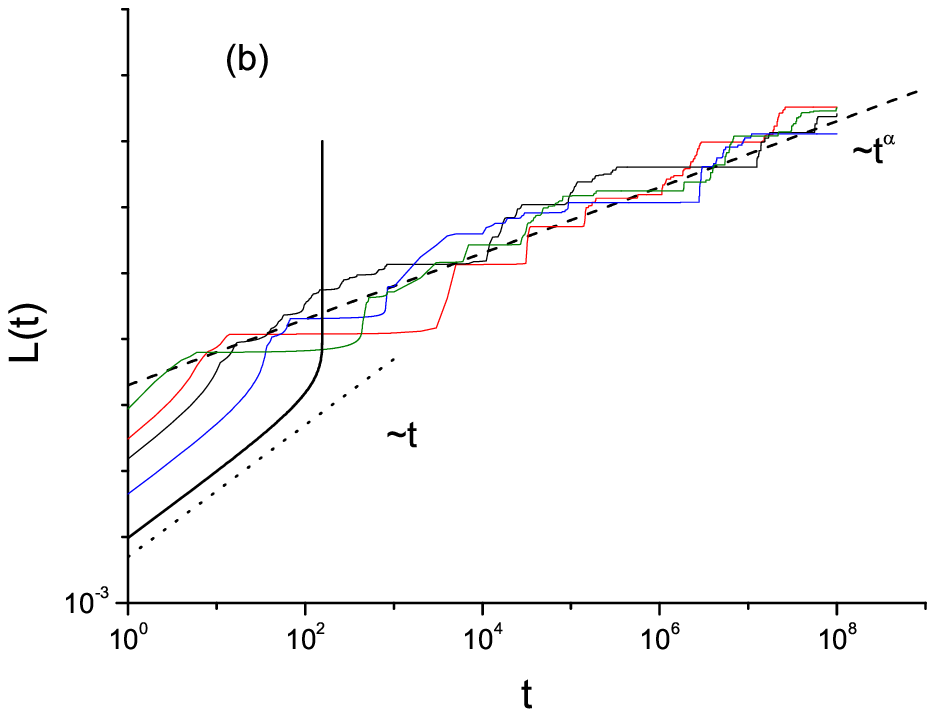,width=80mm,height=60mm}}
\caption{(color online) Function $L(t)=\sum_{i=0}^{t-1} \ln \left| M'(x_i) \right|$ calculated for different single trajectories of the Pomeau-Manneville map Eq.\ (\ref{map_eq_1}) with (a) $z = 1.7$, $a = 1$ and (b) $z = 3$, $a = 1$. For later case notice the small time linear behavior (the dotted line is proportional to $t$) and the long time average power law asymptotic (the dashed line is proportional to $t^{\alpha}$). Divergent line represents $\ln \zeta(t)$ with $\zeta(t)$ given by Eq.\ (\ref{sL}) (see the text for more details).}
\label{Fig_L}
\end{figure}

For a one-dimensional ergodic dynamical system $x_{t+1}=M(x_t)$ the Lyapunov exponent is defined as \cite{ER,Dor}
\begin{equation}
\label{Lyap}
\lambda = \lim_{t\rightarrow \infty} \frac{1}{t} \sum_{i=0}^{t-1} \ln \left| M'(x_i) \right| = \int dx \; \rho(x) \ln \left| M'(x) \right|,
\end{equation}
where $\rho(x)$ is the invariant density of the system and $M'(x)$ is the spatial derivative of the map. By ergodicity
the time average is equal to the average over the invariant density. The existence of the finite limit implies that $L(t) = \ln \left| dx_t/dx_0 \right| = \sum_{i=0}^{t-1} \ln \left| M'(x_i) \right| \propto t$. This behavior is observed for $z<2$ ($\alpha=1$) as shown in Fig.\ \ref{Fig_L} (a). Functions $L(t)$ calculated for different initial conditions collapse on a single asymptotic with the slope given by the Lyapunov exponent $\lambda$.  

For $z>2$ ($0<\alpha<1$) the behavior of $\sum_{i=0}^{t-1} \ln \left| M'(x_i) \right|$ is very different (see Fig.\ \ref{Fig_L} (b)). The density function of trajectories generated by maps Eqs.\  (\ref{map_eq_1},\ref{map_eq_2}) is concentrated on unstable fixed points in the long time limit. The derivative $\left| M'(x) \right|$ at these points is equal to $1$, so $\ln \left| M'(x) \right| \sim \ln 1 \sim 0$, $\lambda=0$ as shown already in \cite{GW88}. Such a behavior is found since most of the time the particle spends in the vicinity of the marginally stable fixed points. In this case the separation of trajectories is stretched exponential \cite{GW88} (see Fig.\ \ref{Fig_L} (b))
\begin{equation}
\label{Lyap_1}
dx_t \sim dx_0 \; e^{\lambda_{\alpha}t^{\alpha}}.
\end{equation}
For different initial conditions $L(t)$ is a random function. The long time behavior of the averaged over initial conditions $L(t)$ is given by $\left< L(t) \right> = \left< \sum_{i=0}^{t-1} \ln \left| M'(x_i) \right| \right> \propto t^{\alpha}$ with $0<\alpha<1$. 

Using the chain rule and the dynamical mapping $x_{t+1} = M(x_t)$, we define
\begin{equation}
\label{scaledFTLE}
\lambda_{\alpha}(x_0) = \frac{1}{t^{\alpha}} \sum_{i=0}^{t-1} \ln \left| M'(x_i) \right|,
\end{equation}
where the dependence on initial condition is emphasized. For $z>2$ ($0<\alpha<1$) $\lambda_{\alpha}$ does not converge to a constant, rather it is a random variable. To prove that a system actually exhibits stretched exponential separation of trajectories, it is sufficient to find the limit distribution of $\lambda_{\alpha}(x_0)$ and to show that it is not trivial (i.e. not a delta function). In Sec. \ref{FTLE_pdf} we will calculate the distribution of $\lambda_{\alpha}$ and will use it to define the generalization of the Lyapunov exponent. However, first in section \ref{PDF} we will investigate infinite invariant densities of the considered maps.

\subsection{Short time behavior}

Here we investigate the short time linear behavior of $L(t) \propto t$  (Fig.\ \ref{Fig_L} (b)). To explain this linear behavior we consider the particles starting from an initial condition very close to $x_t = 0$ and make the continuous time approximation of Eq. (\ref{map_eq_1}) \cite{HHS,Grigo1}, $dx_t/dt = a x_{t}^{z}$ with $x_t \ll 1$. Using this approximation the separation of trajectories close to unstable fixed point is found to be \cite{Grigo1}
\begin{equation}
\label{sL}
\zeta(t)=dx_t/dx_0=\left[1 - (z-1) x_{0}^{z-1} t \right]^{-z/(z-1)},
\end{equation}
where $x_0$ is the initial condition. $\ln \zeta(t)$ is shown in Fig.\ \ref{Fig_L} (b) (divergent line). The short times behavior of $\ln \zeta(t)$ is given by linear in time law
\begin{equation}
\label{ssL}
\ln \zeta(t) \sim z x_{0}^{z-1} t.
\end{equation}
Note that Eq.\ (\ref{sL}) is valid for $t_c<\left[(z-1) x_{0}^{z-1}\right]^{-1}$ since $\ln \zeta(t)$ diverges at $t=t_c$ and is complex for $t>t_c$. Moreover Eq.\ (\ref{sL}) is derived only for particles starting at $t=0$ from an initial condition close to $0$, namely $x_0 \ll 1$, which is also the condition for continuous time approximation to hold. 

The observed linear separation of trajectories is related to aging \cite{Barkai03} in the following sense: if we 'age' the system for time $t_a$ (that is let the system evolve without measurements) and calculate the separation of trajectories after $t_a$, we will observe linear in time behavior for $t \ll t_a$ due to the increased probability for a particle to be in the vicinity of the marginally stable fixed points after aging.


\section{Invariant density}\label{PDF} 

\begin{figure}[tbp]
\centerline{\psfig{figure=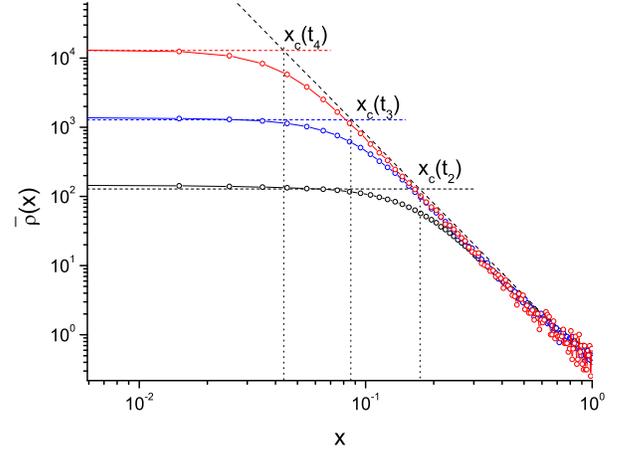,width=90mm,height=70mm}}
\caption{(color online) Numerically calculated invariant density function for the map Eq.\ (\ref{map_eq_1}) with $\alpha=0.3$. Uniform initial density was used. Curves corresponds to different computation times $t_i=10^i$, from bottom to top $i=2,3,4$. Dashed lines correspond to Eq.\ (\ref{g_inv}) with no fitting parameters. The $x_c$ represents the crossover from one asymptotic to another. Note, that $x_c$ decreases with time and $\lim_{t \rightarrow \infty} x_c(t)=0$. For $x \gg x_c$ $\bar{\rho}(x) \sim x^{-1/\alpha}$ yields an infinite invariant density.}
\label{Fig_inv_den}
\end{figure}

First consider model 1. To obtain its invariant density analytically we use the approximation of the map \cite{HHS,Grigo1} for $x\ll1$, $dx_t/dt \simeq a x_{t}^{z}$, and extend it to be valid on the interval $(0,\xi)$ where $\xi$ is defined as the solution of the equation $M(\xi)=1$ (see Fig.\ \ref{Fig_Manneville_map} (a)). When the trajectory reaches the boundary $x=\xi$ it is randomly reinjected back to the interval $(0,\xi)$. The density function $\rho_c(x,t)$ of this system is governed by the equation \cite{HHS,Grigo1}
\begin{equation}
\label{g_Man}
\frac{\partial \rho_c(x,t)}{\partial t} = - \frac{\partial}{\partial x}(a x^z \rho_c(x,t)) + a \xi^z \rho_c(\xi,t),
\end{equation}
where $\xi$ is teated as a parameter. The subscript $c$ in $\rho_c$ is for continuous approximation. The first term on the RHS of Eq.\ (\ref{g_Man}) represents deterministic escape from the marginally stable fixed point, the second term accounts for reinjection of particles. The solution of this equation in Laplace space is
\begin{equation}
\label{g}
\tilde{\rho}_c(x,s) = \frac{\xi^{-1}\tilde{O}_x(s)}{1-a \xi^{z-1}\tilde{O}_{\xi}(s)}, 
\end{equation}
where
\begin{equation}
\label{O_s}
\tilde{O}_x(s)= b(z-1) \left[1- (bs)^{\frac{1}{z-1}}\Gamma\left(\frac{z-2}{z-1},bs\right)\right], 
\end{equation}
and $b=(z-1)^{-1} \; a^{-1} x^{1-z}$. For $0 < \alpha < 1$ ($z>2$) the small $s$ behavior (equivalent to $t \rightarrow \infty$) is given by 
\begin{equation}
\label{g_s} 
\tilde{\rho}_c(x,s) \sim \frac{a^{\alpha-1} x^{-\frac{1}{\alpha}}}{\alpha^{\alpha}\Gamma(1-\alpha)} \; \frac{1}{s^\alpha}, 
\end{equation}
Since, $bs=\alpha a^{-1} x^{-1/\alpha} s$ is the small parameter of the expansion, Eq.\ (\ref{g_s}) is valid for all $x \ne 0$. For $x = 0$ we get different expression
\begin{equation}
\label{g_s_1} 
\tilde{\rho}_c(0,s) \sim \frac{1}{\alpha^{\alpha}\Gamma(1-\alpha)} \; \frac{1}{s^{1+\alpha}}. 
\end{equation}
In practice there is a crossover from one asymptotic to another at some $x_c \ll 1$ which we define below. Transforming Eqs.\ (\ref{g_s}, \ref{g_s_1}) into the time domain and considering $t \rightarrow \infty$ one gets for $0 < \alpha < 1$ ($z>2$)
\begin{equation}
\label{g_t} 
\rho_c(x,t) \sim \begin{cases}  \frac{a^{\alpha-1} x^{-\frac{1}{\alpha}}}{\alpha^{\alpha}} \; \frac{\sin(\pi \alpha)}{\pi} \; t^{\alpha-1}, & x \gg x_c \cr
\frac{\sin(\pi \alpha)}{\pi \alpha^{1+\alpha}}\; t^{\alpha}, & x \ll x_c.
\end{cases}
\end{equation}
We define the crossover $x_c$ as $\rho_{c \; (x \gg x_c)}(x_c)=\rho_{c \; (x \ll x_c)}(x_c)$ (see Fig.\ \ref{Fig_inv_den}). Using Eq.\ (\ref{g_t}), the time dependence of the crossover is obtained $x_c=\alpha^{\alpha} t^{-\alpha}$. Hence, the crossover goes to zero $x_c \rightarrow 0$ as $t \rightarrow \infty$. For $z<2$ we find the following solution for the density function
\begin{equation}
\label{g_s2}
\tilde{\rho}_c(x,s) \sim \begin{cases}  (2-z) \; x^{1-z} \frac{1}{s}, & x \gg x_c \cr
(2-z) \; \frac{1}{s^2}, & x \ll x_c,
\end{cases}
\end{equation}
and
\begin{equation}
\label{g_t2}
\rho_c(x,t) \sim \begin{cases} (2-z) \; x^{1-z}, & x \gg x_c \cr
(2-z) \; t, & x \ll x_c.
\end{cases}
\end{equation}
For $z=2$ the solution is given by
\begin{equation}
\label{g_s3}
\tilde{\rho}_c(x,s) \sim \begin{cases}  \frac{x^{-1}}{s \ln(\frac{1}{s})}, & x \gg x_c \cr
\frac{1}{s^2 \ln(\frac{1}{s})}, & x \ll x_c,
\end{cases}
\end{equation}
and
\begin{equation}
\label{g_t3}
\rho_c(x,t) \sim \begin{cases} \frac{x^{-1}}{\ln(t)}, & x \gg x_c \cr
\frac{t}{\ln(t)}, & x \ll x_c.
\end{cases}
\end{equation}
Note, that the density function is stationary only for $z<2$ and $x \gg x_c$ (Eq.\ (\ref{g_t2})). For $z\ge2$ the density function depends on time even in the long time limit. Moreover, the dependence is of the power law form for $0 < \alpha < 1$ (Eq.\ (\ref{g_t})) while it is logarithmic for $z=2$ (Eq.\ (\ref{g_t3})). We introduce  
\begin{equation}
\label{g_inv}
\bar{\rho}_c(x,t) = t^{1-\alpha} \; \rho_c(x,t) = \begin{cases}  \frac{a^{\alpha-1}}{\alpha^{\alpha}} \; \frac{\sin(\pi \alpha)}{\pi} \; x^{-\frac{1}{\alpha}}, & x \gg x_c \cr
\frac{\sin(\pi \alpha)}{\pi \alpha^{1+\alpha}}\; t, & x \ll x_c,
\end{cases}
\end{equation}
for $0 < \alpha < 1$. For $z=2$ ($\alpha=1$) the scaling factor is $\ln(t)$
\begin{equation}
\label{g_inv_3}
\bar{\rho}_c(x,t) = \ln(t) \; \rho_c(x,t) = \begin{cases}  x^{-1}, & x \gg x_c \cr
t, & x \ll x_c.
\end{cases}
\end{equation}
Note that in the long time limit $x_c\rightarrow0$ and scaled density functions become independent of time
$\bar{\rho}_c(x,t) \rightarrow \bar{\rho}_c(x)$  
\begin{equation}
\label{g_inv_4}
\bar{\rho}_c(x) = \begin{cases} \frac{a^{\alpha-1}}{\alpha^{\alpha}} \; \frac{\sin(\pi \alpha)}{\pi} \; x^{-\frac{1}{\alpha}}, & 0 < \alpha < 1, \cr
x^{-1}, & \alpha=1.
\end{cases}
\end{equation}
For all $\alpha$ the scaled densities have power law decay in $x$, $\bar{\rho}_c(x)\sim x^{-1/\alpha}$, and their integrals diverge for $0 < \alpha \le 1$ (when we take the long time limit $x_c=0$)
\begin{equation}
\int_{0}^{\xi} dx \bar{\rho}_c(x) = \infty.
\end{equation}
Thus, $\rho_c(x)$ is not normalizable. Still, as we show later, the infinite invariant density is useful for the calculation of the statistical properties of the dynamics. 
\begin{figure}[tbp]
\centerline{\psfig{figure=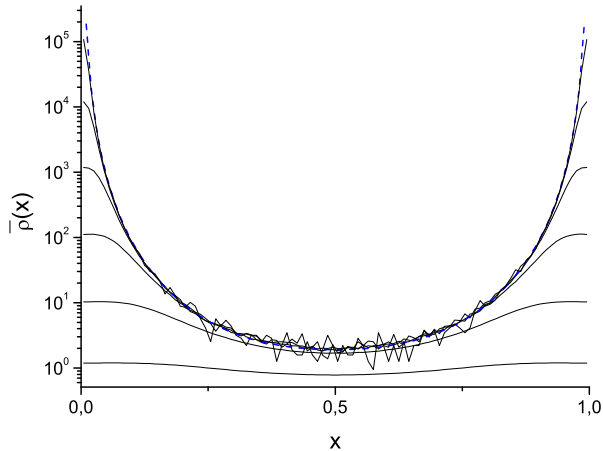,width=90mm,height=70mm}}
\caption{(color online) The same as in Fig.\ \ref{Fig_inv_den} but for the map Eq.\ (\ref{map_eq_2}) with $\alpha=0.3$. Curves corresponds to different computation times $t=10^i$ from bottom to top $i=0,1,2,3,4,5$. Dashed line is the best fit (see Eq.\ (\ref{rho_approx})). We have $\bar{\rho}(x) \sim x^{-1/\alpha}$ for $x \rightarrow 0$ and $\bar{\rho}(x) \sim (1-x)^{-1/\alpha}$ for $x \rightarrow 1$. Thus, non normalizable infinite invariant density emerges.}
\label{Fig1}
\end{figure}

We perform numerical simulations and obtain the invariant density of Eq.\ (\ref{map_eq_1}) comparing it with the density function obtained analytically. In these simulations we start with a uniform density and plot $\bar{\rho}(x)=t^{1-\alpha}\rho(x,t)$ versus $x$. Results are shown in Fig.\ \ref{Fig_inv_den}. The analytical solutions of the continuous stochastic model Eq.\ (\ref{g_inv}) are plotted as dashed lines. We find excellent agreement between continuous model Eq.\ (\ref{g_Man}) and numerics without fitting. Horizontal lines represent asymptotic solution for $x \ll x_c$ calculated for the corresponding time of the simulation, while the sloping line corresponds to the asymptotic solution for $x \gg x_c$ which decays as $x^{-\frac{1}{\alpha}}$. The $x_c$ represents the crossover from one asymptotic to another. As $t \rightarrow \infty$, $x_c \rightarrow 0$ and we approach the infinite invariant density.

For the second model Eq.\ (\ref{map_eq_2}) with two marginally unstable fixed points an example of numerically calculated invariant density function for $\alpha=0.3$ is shown in Fig.\ \ref{Fig1}. The invariant density function in this case is well approximated by
\begin{equation}
\label{rho_approx} 
\bar{\rho}(x) \sim a \; x^{-\frac{1}{\alpha}} + a \; (1-x)^{-\frac{1}{\alpha}} + b,
\end{equation}
where $a=0.04$ and $b=1$ for $\alpha=0.3$. In this case the infinite invariant density has two peaks on $x = 0$ and $x = 1$ corresponding to two unstable fixed points. As before $\int_0^1 dx \; \bar{\rho}(x)=\infty$.


\section{Generalized Lyapunov exponent}\label{FTLE_pdf}
  
\begin{figure}[tbp]
\centerline{\psfig{figure=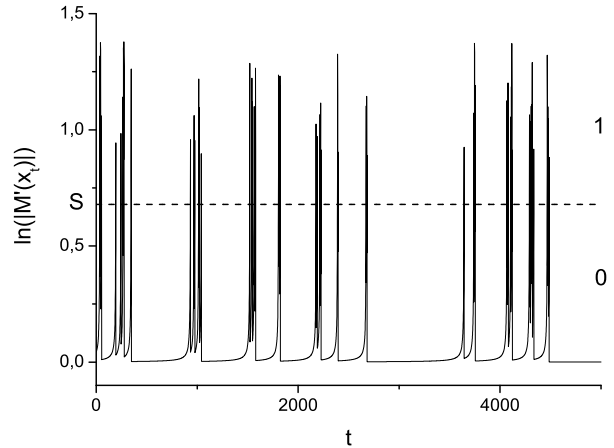,width=90mm,height=70mm}}
\caption{Illustration of the renewal process $I(t)$ defined by $\ln(\left|M'(x_t)\right|)$ for the map Eq.\ (\ref{map_eq_1}) with $\alpha=0.3$, $a=1$. The dashed line is a threshold $S$ which is chosen to be the discontinuity point of the map Eq.\ (\ref{map_eq_1}), $S=\xi$.}
\label{Manneville_trac}
\end{figure}

Now we return to the investigation of distribution of $\lambda_{\alpha}$ for $z>2$ ($0<\alpha<1$). This problem can be treated using the AKD theorem \cite{TZ06,Thaler02,Aaronson}, as pointed out in \cite{JKP}, however we feel that a rederivation using stochastic arguments is useful. We consider the logarithm of the derivative of the map $y=\ln \left| M'(x_t) \right|$ illustrated in Fig.\ \ref{Manneville_trac} for a single initial condition. Then we define a two state process $I(t)=0$ if $y<S$ and $I(t)=1$ if $y>S$, where $S$ denotes a threshold. Waiting times in state $0$ are distributed according to power laws with an infinite sojourn time \cite{GT84,Zum93} (see details below), while waiting times in state $1$ have a characteristic average time. Ignoring therefore times spent in state $1$ and neglecting correlations we consider $I(t)$ as a renewal process. Let $n$ be the number of renewals which occur in time $t$, namely, number of transitions from state $0$ to $1$. The logarithm of the derivative of the map $\ln \left| M'(x_t) \right|$ is equal to zero most of the time since the trajectory stays for long times near marginally unstable fixed points, only for short periods its value deviates from zero (see Fig.\ \ref{Manneville_trac}). The sum of logarithms along the trajectory is thus proportional to $n$: $\sum_{i=0}^{t-1} \ln \left| M'(x_i) \right| \propto c \; n$, where $c$ is a constant, which according to the renewal theory \cite{Feller} is a random variable. Our goal is to calculate the pdf of $\lambda_{\alpha}$ Eq.\ (\ref{scaledFTLE}). Note that the distribution of 'standard' local Lyapunov exponent $\lambda= \sum_{i=0}^{t-1} \ln \left| M'(x_i) \right|/t$ (where $t$ is finite) was investigated in \cite{TA93}. This probability density function (PDF) was found to consist of {\bf a} smooth part that vanish as $t \rightarrow \infty$ and a peak at $x=0$ (corresponding to $\lambda=0$ as $t \rightarrow \infty$).
\begin{figure}[tbp]
\centerline{\psfig{figure=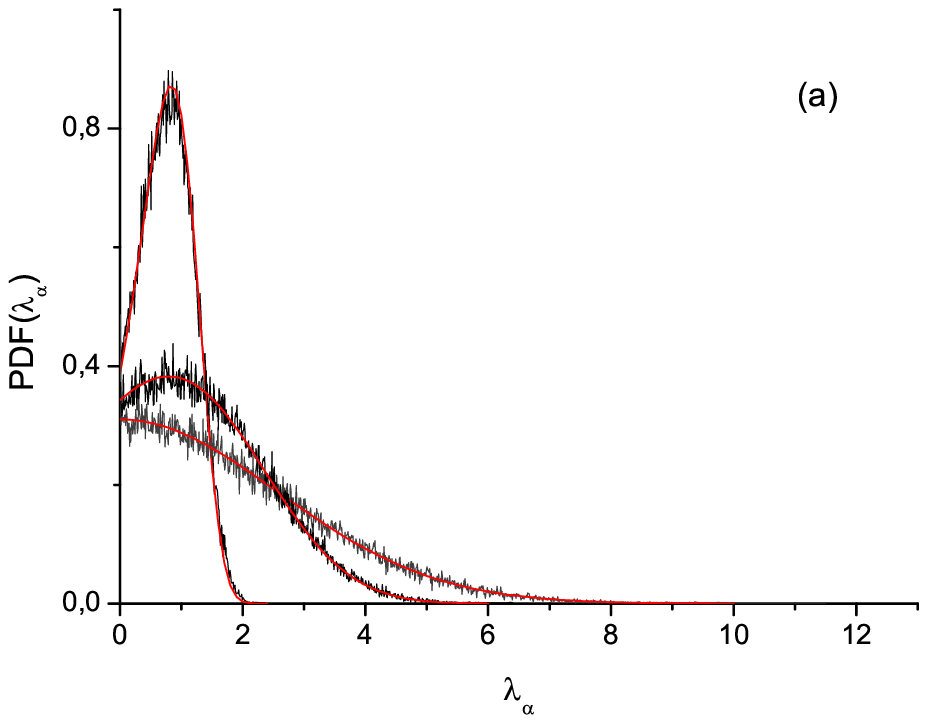,width=90mm,height=70mm}}
\centerline{\psfig{figure=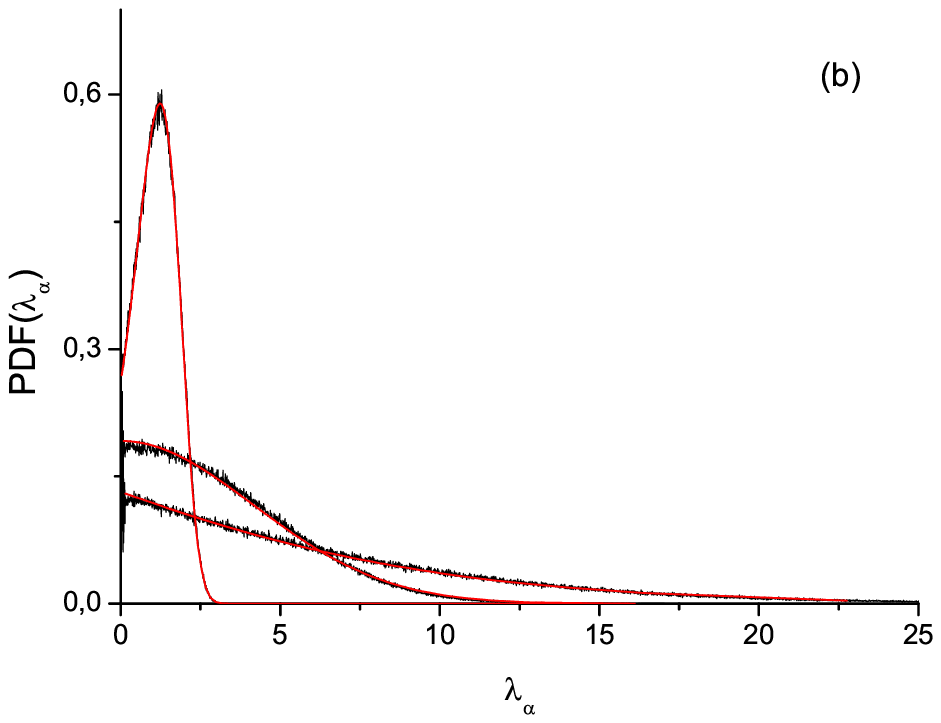,width=90mm,height=70mm}}
\caption{(color online) (a) Numerically calculated PDF of $\lambda_{\alpha}$ for the map Eq.\ (\ref{map_eq_1}) with $\alpha = 0.75,0.59,0.5$ calculated with $10^5$ trajectories iterated for $t = 10^5$. Smooth curves (which coincide with numerical data without fitting) correspond to analytical PDF Eq.\ (\ref{pdfL_zeta}) with $\left< \lambda_{\alpha} \right>$ calculated with Eq.\ (\ref{L_mean_c}). (b) The same as in (a) but for the map Eq.\ (\ref{map_eq_2}) with $\alpha=0.75, 0.5, 0.3$ from top to bottom calculated with $10^6$ trajectories iterated for $t = 10^4$ for $\alpha = 0.75, 0.5$ and $t = 10^5$ for $\alpha = 0.3$. Smooth curves correspond to analytical PDF Eq.\ (\ref{pdfL_zeta}) with $\left< \lambda_{\alpha} \right>$ calculated numerically using Eq.\ (\ref{L_mean_3}).}  
\label{Manneville_PDF_L}
\end{figure}

The probability to have exactly $n$ renewal events in time $t$ can be expressed through the waiting time distribution as well known \cite{Feller}. It is calculated as follows: the PDF of time intervals between renewals can be obtained from the continuous approximation of the maps near marginally unstable fixed points $dx_t/dt \simeq a x^z$. Solving this equation one finds the duration of laminar motion as function of initial condition $x_0$ \cite{GT84}
\begin{equation}
\label{xt_x0}
t = \frac{a}{\alpha} \left[ x_{0}^{-\frac{1}{\alpha}} - 1 \right].
\end{equation}
The PDF $\psi(t)$ of the waiting times is thus related to the distribution $P_{in}(x_0)$ of
injection points which we assume to be uniform $P_{in}(x_0)=1$, $\psi(t)dt \equiv P_{in}(x_0)dx_0$, implying \cite{GT84}
\begin{equation}
\label{waiting_pdf}
\psi(t)\sim A/t^{1+\alpha}, \; \; t \rightarrow \infty,
\end{equation}
\begin{equation}
\label{waiting_pdf_laplace}
\tilde{\psi}(s) = \int_{0}^{\infty} dt \; e^{-st} \psi(t) \simeq 1 - B s^{\alpha}, \; s \rightarrow 0,
\end{equation}
where $\tilde{\psi}(s)$ is the Laplace transform of $\psi(t)$ and $A$, $B$ are some positive constants. Let $P_n(t)$ be the probability of having $n$ renewal events in $(0,t)$ and $P_n(s)$ it Laplace transform. Using convolution theorem \cite{Feller}
\begin{equation}
\tilde{P}_n(s) = \frac{\tilde{\psi}^n(s)\left(1-\tilde{\psi}(s)\right)}{s}.
\end{equation}
For small $s$ corresponding to large $t$ one obtains
\begin{equation}
\label{rpdfL}
\tilde{P}_n(s) \simeq \frac{Be^{-Bns^{\alpha}}}{s^{1-\alpha}}.
\end{equation}
The inverse Laplace transform of Eq.\ (\ref{rpdfL}) can be expressed in terms of one-sided L{\'e}vy PDF \cite{Feller}
\begin{equation}
\label{renewal_pdf}
P_n(t) = \frac{1}{\alpha} \frac{t}{n^{1+1/\alpha}B^{1/\alpha}} l_{\alpha}\left[\frac{t}{\left( Bn\right)^{1/\alpha}}\right],
\end{equation}
where the one-sided L{\'e}vy PDF is defined through its Laplace transform $\tilde{l}_{\alpha}(s) = exp(-s^{\alpha})$. 

Since for the calculation of the distribution of $\lambda_{\alpha}$ (Eq.\ (\ref{scaledFTLE})) one needs to have the unknown constant $c$, we proceed to the rescaled variable $\zeta=\lambda_{\alpha}/\left< \lambda_{\alpha} \right>$. Here $\left< \lambda_{\alpha} \right>$ is the mean value of $\lambda_{\alpha}$ which we calculate later. According to the renewal assumption, $\zeta=n/\left< n \right>$. Using Eq.\ (\ref{renewal_pdf}) the mean number of renewal events is given by $\left< n \right> = \frac{t^{\alpha}}{B \Gamma(1+\alpha)}$. By change of variables we arrive at the PDF of rescaled variable $\zeta$
\begin{equation}
\label{renewal_pdf_zeta}
P_{\alpha}(\zeta) = \frac{\Gamma^{1/\alpha}(1+\alpha)}{\alpha \zeta^{1+1/\alpha}} \; l_{\alpha}\left[\frac{\Gamma^{1/\alpha}(1+\alpha)}{\zeta^{1/\alpha}}\right].
\end{equation}
This is the first main result of the manuscript which is a special case of the more general ADK theorem \cite{TZ06,Thaler02,Aaronson}. This result was obtained previously in Ref.\ \cite{JKP}, as mentioned. One can also express Eq.\ (\ref{renewal_pdf_zeta}) in terms of the Mittag-Leffler PDF \cite{Feller,BM97}. To get the distribution of $\lambda_{\alpha}$ we change variables again, using $\zeta=\lambda_{\alpha}/\left<\lambda_{\alpha}\right>$ and Eq.\ (\ref{renewal_pdf_zeta})
\begin{equation}
\label{pdfL_zeta}
P_{\alpha}(\lambda_{\alpha}) = \frac{1}{\left< \lambda_{\alpha} \right>} P_{\alpha}(\zeta).
\end{equation}
By definition $\left< \lambda_{\alpha} \right>$ is given by
\begin{equation}
\label{lambda-bar}
\left< \lambda_{\alpha} \right> = \int_{0}^{\infty} \lambda_{\alpha} \; P_{\alpha}(\lambda_{\alpha}) \; d\lambda_{\alpha}.
\end{equation}
For $z<2$ we have $\alpha=1$ (see Eq.\ (\ref{alf})) and the distribution of finite time Lyapunov exponent is the delta-function $P_1(\lambda_1)=\delta(\lambda_{1} - \lambda)$ and $\left< \lambda_{1} \right>$ reduces to the standard Lyapunov exponent Eq.\ (\ref{Lyap}).
\begin{figure}[t]
\centerline{\psfig{figure=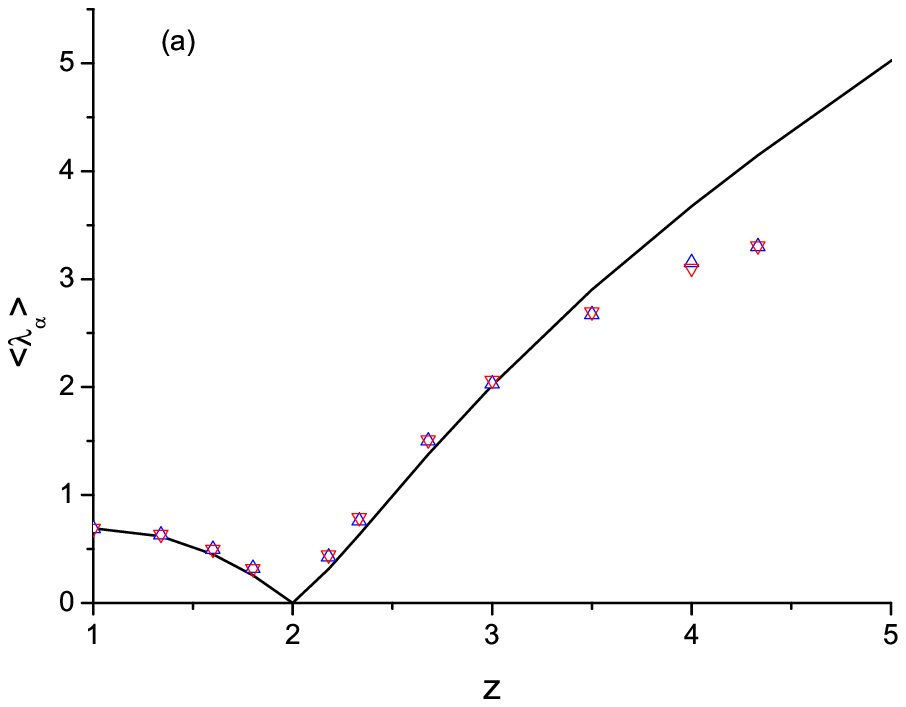,width=90mm,height=70mm}}
\centerline{\psfig{figure=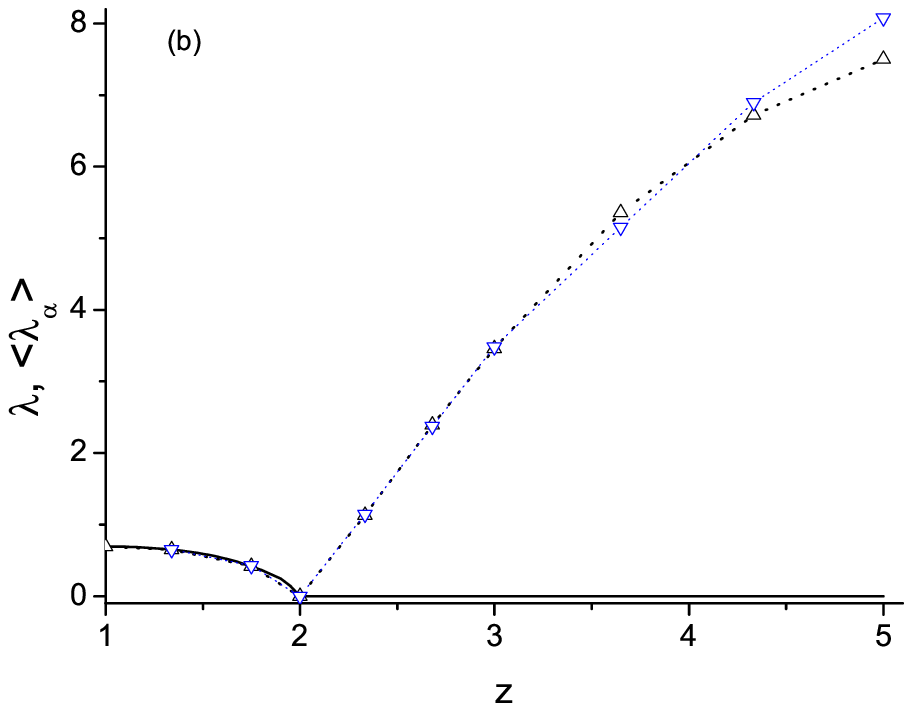,width=90mm,height=70mm}}
\caption{(color online) (a) Symbols represent $\left< \lambda_{\alpha} \right>$ calculated by Eqs.\ (\ref{lambda-bar},\ref{L_mean_3}) for the map Eq.\ (\ref{map_eq_1}). Solid line is $\left< \lambda_{\alpha} \right>$ calculated with analytical infinite invariant density Eq.\ (\ref{L_mean_c}). (b) Symbols represent the same as on the top panel but for the map Eq.\ (\ref{map_eq_2}). Dashed line is the standard Lyapunov exponent $\lambda$ Eq.\ (\ref{Lyap}).}
\label{Manneville_GLE}
\end{figure}

Now we calculate the average $\left< \lambda_{\alpha} \right>$, which is the main new result of this section. 
Using Eq.\ (\ref{scaledFTLE})
\begin{equation}
\label{L_mean}
\left< \lambda_{\alpha} \right> = \int_0^1 \frac{\sum_{i=0}^{t-1} \ln \left| M'(x_i) \right|}{t^{\alpha}} \; \rho(x_0) \; dx_0,
\end{equation}
where the averaging is over initial conditions distributed according to some initial density $\rho(x_0)$. Since we are interested in the long time limit we replace the summation with an integral and average over the density function 
\begin{equation}
\label{L_mean_1}
\left< \lambda_{\alpha} \right> \simeq \frac{1}{t^{\alpha}} \int_0^1 dx \int_{0}^{t} \ln \left| M'(x) \right|\; \rho(x,t') \; dt'.
\end{equation}
For Pomeau-Manneville map Eq.\ (\ref{map_eq_1}) the density function has two asymptotics valid for $x \ll x_c$ and $x \gg x_c$ Eq.\ (\ref{g_t}). Hence,
\[
\left< \lambda_{\alpha} \right> \simeq \frac{1}{t^{\alpha}} \int_0^{x_c} dx \int_{0}^{t} \ln \left| M'(x) \right|\; \rho_{(x \ll x_c)}(x,t') \; dt' +
\]
\begin{equation}
\label{L_mean_2}
+ \frac{1}{t^{\alpha}} \int_{x_c}^1 dx \int_{0}^{t} \ln \left| M'(x) \right|\; \rho_{(x \gg x_c)}(x,t') \; dt',
\end{equation}
where $x_c$ denotes the crossover from one asymptotic of $\rho(x,t)$ to another (see Fig.\ \ref{Fig_inv_den}). The crossover $x_c$ was defined in the previous section as $\rho_{(x \gg x_c)}(x_c)=\rho_{(x \ll x_c)}(x_c)$. Using Eq.\ (\ref{g_t}), $x_c = \alpha^{\alpha} t^{-\alpha}$ and as $t \rightarrow \infty$, $x_c \rightarrow 0$. As a result the first integral vanishes in the long time limit. The fact that this integral vanishes is not trivial since $\rho_{(x \ll x_c)}\rightarrow \infty$ (see Eq.\ (\ref{g_t})), it happens because we consider a specific observable with $\ln \left| M'(x) \right| \rightarrow 0$ as $x\rightarrow 0$, which cancels the $t^{\alpha}$ divergence found in Eq.\ (\ref{g_t}). Using the density function in the form $\rho(x,t)\simeq C \; t^{\alpha-1}/x^{1/\alpha}$ for $x \gg x_c$ (Eq.\ (\ref{g_t})), we arrive at 
\begin{equation}
\left< \lambda_{\alpha} \right> \simeq \frac{1}{t^{\alpha}} \int_0^1 dx \int_{0}^{t} \ln \left| M'(x) \right|\;  \frac{C \; \tau^{\alpha-1}}{x^{1/\alpha}} \; d\tau,
\end{equation}
where $C$ is a positive constant. For the analytical approximation of the map studied in Sec.\ \ref{PDF} constant $C$ is given by $C=\frac{a^{\alpha-1} \sin (\pi \alpha)}{\pi \alpha^{\alpha}}$ (Eq.\ (\ref{g_t})). Finally, computing the integral over time and noticing that $t^{1-\alpha}\rho(x,t)=\bar{\rho}(x)$ (Eq.\ (\ref{g_inv})), we obtain the final result 
\begin{equation}
\label{L_mean_3}
\left< \lambda_{\alpha} \right> = \frac{1}{\alpha} \int_0^1 dx \ln \left| M'(x) \right|\; \bar{\rho}(x).
\end{equation}
Since $\ln \left| M'(x) \right|$ vanishes precisely where the invariant density has divergences, the integral is finite and positive in spite of the fact that the integral of $\bar{\rho}(x)$ diverges. Eq.\ (\ref{pdfL_zeta}) together with Eq.\ (\ref{L_mean_3}) fully characterize the distribution of generalized Lyapunov exponents in the non ergodic phase. 
Eq.\ (\ref{L_mean_3}) is our second main result, and we claim that it is valid for systems with unstable fixed points and not limited to Pomeau-Manneville map. For the stochastic model Eq.\ (\ref{g_Man}) with $0<\alpha<1$ using Eq.\ (\ref{g_inv}) one gets for the map in Eq.\ (\ref{map_eq_1})
\begin{equation}
\label{L_mean_c}
\left< \lambda_{\alpha} \right> = \frac{1}{\alpha} \int_0^1 dx \; \frac{a^{\alpha-1}}{\alpha^{\alpha}} \frac{\sin(\pi \alpha)}{\pi} \; \frac{\ln (1+ a z x^{z-1})}{x^{1/\alpha}} .
\end{equation}
When $z \rightarrow 2$ (i.e. $\alpha \rightarrow 1$) $\left< \lambda_{\alpha} \right>=0$ since then we approach the normal phase, for $z \rightarrow \infty$ (i.e. $\alpha \rightarrow 0$) $\left< \lambda_{\alpha} \right> \sim z$.

Distributions of $\lambda_{\alpha}$ obtained by simulation of maps Eqs.\ (\ref{map_eq_1},\ref{map_eq_2}) are shown in Fig. \ref{Manneville_PDF_L}. Smooth curves correspond to analytical PDF Eq.\ (\ref{pdfL_zeta}) with $\left< \lambda_{\alpha} \right>$ calculated numerically according to Eq.\ (\ref{L_mean_3}). The perfect agreement between theory and simulation indicates that the general theory works well for finite time simulations. The figure demonstrates that one-sided L{\'e}vy distributions describe scaled finite time Lyapunov exponent also for maps with two marginally stable fixed points.    

Numerically computed generalized Lyapunov exponent is shown in Fig.\ \ref{Manneville_GLE}. For the map Eq.\ (\ref{map_eq_1}) (Fig.\ \ref{Manneville_GLE} (a)) we find good agreement between $\left< \lambda_{\alpha} \right>$ calculated with the approximate analytical infinite invariant density Eq.\ (\ref{L_mean_c}) and numerically computed one. The convergence of $\left< \lambda_{\alpha} \right>$ is getting worse for $z$ close to $z=2$ because of the slow logarithmic convergence to the invariant density at $z=2$. Also the convergence is getting worse for large $z$ (small $\alpha$). In the ergodic phase $z<2$ the standard Lyapunov exponent Eq.\ (\ref{Lyap}) is recovered. In particular, for $z=1$ the analytical value $\lambda=\ln(2)$ is obtained \cite{Dor}. In the non ergodic phase $z>2$ ($0<\alpha<1$) the generalized Lyapunov exponent is finite while the standard Lyapunov exponent Eq.\ (\ref{Lyap}) vanishes. 

For the second model Eq.\ (\ref{map_eq_2}) the behavior of $\left< \lambda_{\alpha} \right>$ is qualitatively the same (Fig.\ \ref{Manneville_GLE} (b)). Solid line here represents the standard Lyapunov exponent which vanishes for $z>2$. 


\section{Krengel's entropy and Pesin-type Identity}\label{GLEs}

\begin{figure}[t]
\centerline{\psfig{figure=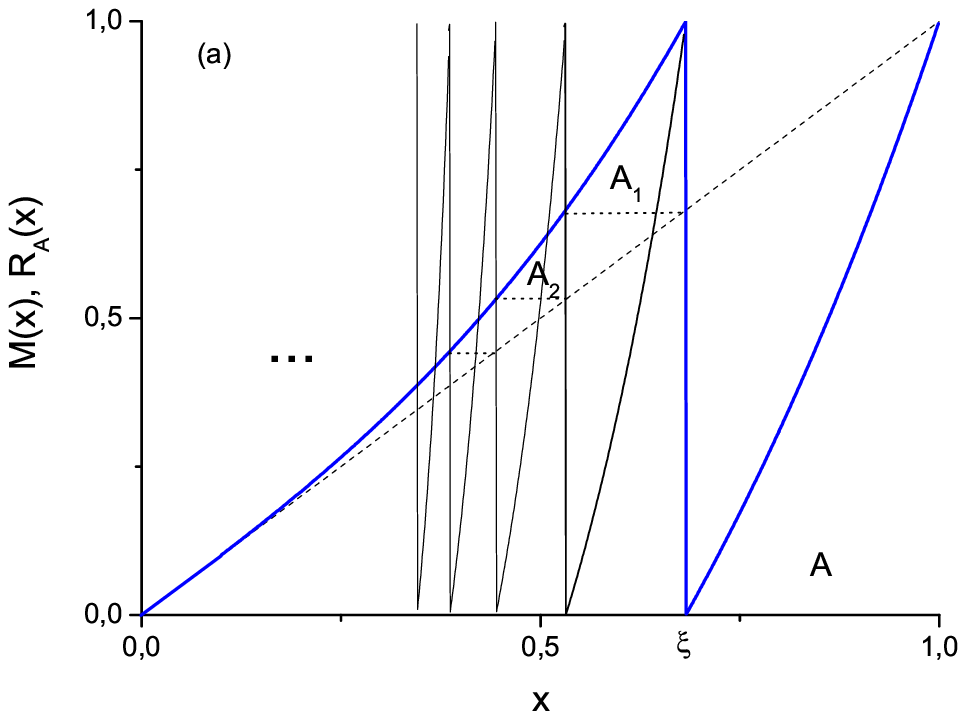,width=90mm,height=70mm}}
\centerline{\psfig{figure=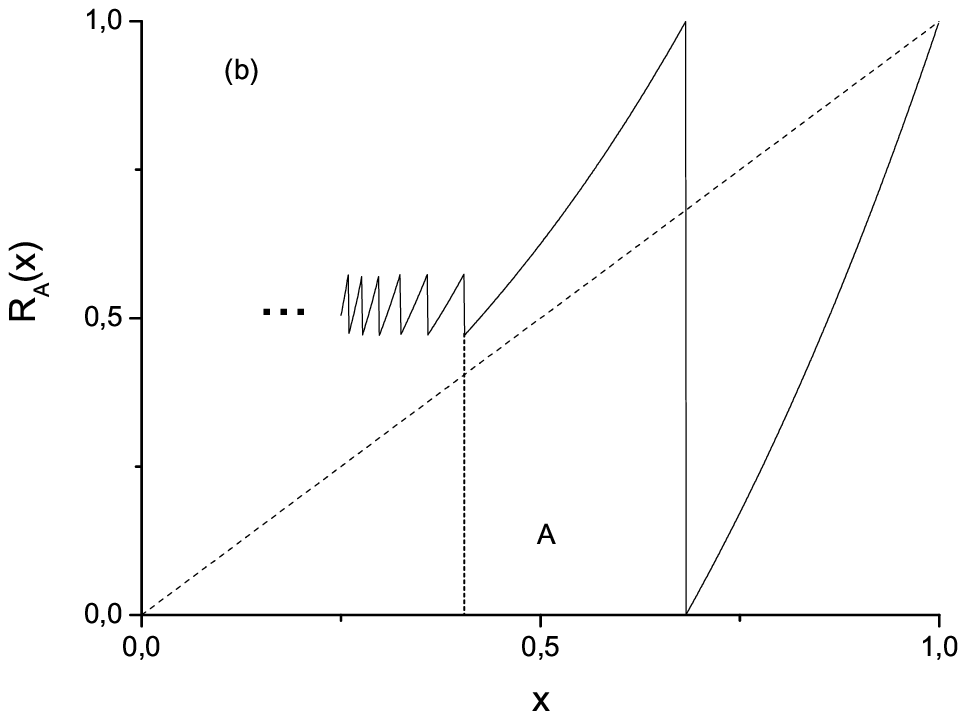,width=90mm,height=70mm}}
\caption{(color online) (a) The map $M(x)$ in Eq.\ (\ref{map_eq_1}) with $z = 3$, $a = 1$ (dashed line) and its first return map (FRM) $R_A(x)$ with the return interval $A = [\xi, 1]$. (b) $R_A(x)$ for $A = [0.47, 1]$.}
\label{Manneville_FRM}
\end{figure}

The entropy $h_{\alpha}$ for infinite measure preserving transformations was introduced by Krengel as the Kolmogorov-Sinai entropy of its first return map (FRM) times the integral over the invariant density \cite{Krengel}
\begin{equation}
\label{Krengel}
h_{\alpha} = h_{KS} (R_A) \int_A dx \bar{\rho}(x).
\end{equation}
The FRM $R_A$ is defined on any subset $A$ of finite measure as $M^{\tau(x)}(x)$, where $\tau(x) = 1 + n(x)$ and $n(x)$ is the smallest positive integer such that $M^n \in A$ \cite{Pia80}. An example of the construction of the FRM for the Manneville map Eq.\ (\ref{map_eq_1}) with $A = [\xi, 1]$ is shown in Fig.\ \ref{Manneville_FRM} (a). All trajectories starting in $A_1$ are mapped into $A$ already in one iteration, so $n(x) = 1$, $\tau(x) = 2$ and $R_A(x) = M(M(x))$.
All $x \in A_2$ are mapped into $A_1$ and after that from $A_1$ to $A$, so $n(x) = 2$, $\tau(x) = 3$ and $R_A(x) = M(M(M(x)))$. The FRM map $R_A(x)$ has infinite number of branches accumulating for $x \rightarrow 0$. The slope of all branches is strictly greater than $1$. Hence, the FRM is everywhere expanding and therefore has finite invariant probability density, positive Lyapunov exponent and Kolmogorov-Sinai entropy. $h_{\alpha}$ does not depend on the choice of $A$ \cite{DKU90}. Fig.\ \ref{Manneville_FRM} (b) shows how the FRM is modified when $A$ is increased to $[0.4, 1]$. Although the integral $\int_A dx \bar{\rho}(x)$ is increased compared to the case $A = [\xi, 1]$, the FRT map and its $h_{KS}(R_A)$ is also changed to compensate this increase.

Krengel's entropy $h_{\alpha}$ satisfies Rohlin's formula \cite{Thaler83,Zwe00,Zwe06,Rohlin}
\begin{equation}
\label{h_alf}
h_{\alpha} = \int \ln \left| M'(x) \right| \bar{\rho}(x) \; dx,
\end{equation}
where $\bar{\rho}(x)$ is the infinite invariant density. Comparing Eq.\ (\ref{h_alf}) with Eq.\ (\ref{L_mean_3}) we find that up to a constant $\alpha$, Krengel's entropy $h_{\alpha}$ is equal to our generalized Lyapunov exponent $\left< \lambda_{\alpha} \right>$ calculated with the invariant density
\begin{equation}
\label{GPesin}
h_{\alpha} = \alpha \left< \lambda_{\alpha} \right>.
\end{equation}
Thus, unlike $h_{KS}$ entropy in the non ergodic phase the Krengel's entropy of the system $h_{\alpha}$ is nonzero and positive. Moreover, up to a factor $\alpha$ it is equal to the Lyapunov exponent that generalizes the Pesin's identity.
\begin{figure}[t]
\centerline{\psfig{figure=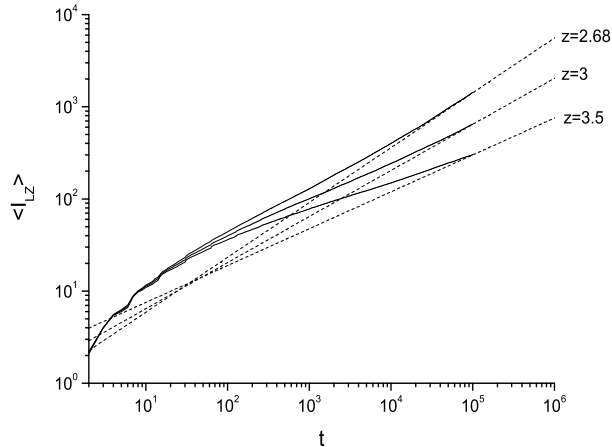,width=90mm,height=70mm}}
\caption{Information content $\left< I_{LZ} \right>$ calculated by the Lempel-Ziv algorithm for the map Eq.\ (\ref{map_eq_1}) with $a = 1$, $z = 2.68$ ($\alpha = 0.59$), $z = 3$ ($\alpha = 0.5$) and $z = 3.5$ ($\alpha = 0.4$) average over $10^3$ trajectories. Dashed lines correspond to $\left< I_{LZ} \right> \sim h_{\alpha} t^{\alpha}$ with $h_{\alpha}$ calculated by Eq.\ (\ref{h_alf}).}
\label{Manneville_LZ}
\end{figure}

What is the Physical meaning of the Krengel's entropy? To answer this question we consider complexity of a single trajectory calculated by Lempel-Ziv algorithm LZ$77$, which originates from the data compression and characterizes degree of randomness in a system \cite{LZ}. Briefly the algorithm can be described as follows: in the first step the signal is transformed into binary sequences. We have used threshold $0 - 1$ coding assigning $'0'$ for $x < \xi$ and $'1'$ for $x > \xi$ (see Fig.\ \ref{Fig_Manneville_map} where $\xi$ is defined for the map Eq.\ (\ref{map_eq_1})). The sequence $S = s_1 s_2 ...$, (where $s_i$ is $'0'$ or $'1'$) is sequentially scanned and rewritten as a concatenation of words $w_1 w_2 ...$ chosen in such a way that $w_1 = s_1$ and a new phrase $w_{k+1}$ is formed by the longest match anywhere in a past, plus the new symbol. In other words, $w_{k+1}$ is the extension of some word $w_j$ in the list, $w_{k+1} = w_j s$, where $0 \le j \le k$, and $s$ is either $0$ or $1$. The corresponding compressed code consists then of pairs of numbers: an indicator being a pointer to the previous occurrence of the prefix of the word and the last bit of the word. As an example consider the string $S$ which is decomposed into words as follows: 
\begin{equation}
\label{S}
S =  00011110101... = (0)(00)(1)(11)(10)(101)... \; .
\end{equation}  
Let $c(t)$ denote the number of words in the sequence. In the above example $c(t)=6$. For each word, we use $\log_2 c(t)$ bits to describe the location of the prefix to the word and $1$ bit to describe the last bit. The information content of the string, $I_{LZ}$, is defined as the total length of the encoded sequence
\begin{equation}
\label{I_LZ}
I_{LZ} = c(t) \left[ \log_2 c(t) + 1 \right].
\end{equation}  
(Note, when $t \rightarrow \infty$ the $1$ is negligible and $I_{LZ} = c \log_2 c$ resembles the Shannon entropy.) The complexity of the sequence for LZ$77$ algorithm is related to the length of the associated encoding via \cite{BP}:
\begin{equation}
\label{C_LZ}
C_{LZ} = \lim_{t \rightarrow \infty} \frac{I_{LZ}}{t}.
\end{equation}  
For a stationary dynamical system with finite invariant measure and positive Kolmogorov-Sinai entropy $h_{KS}$ the complexity of a string is \cite{BP}
\begin{equation}
\label{C_LZ_ergod}
C_{LZ} = \frac{h_{KS}}{\ln 2}.
\end{equation}  
Thus, Eq.\ (\ref{C_LZ}) can be used to estimate the Kolmogorov-Sinai entropy $h_{KS}$ \cite{Gra89}.

Here we apply LZ$77$ algorithm to estimate entropy for intermittent systems. For the Pomeau-Manneville map Eq.\ (\ref{map_eq_1}) the average of the information content was shown to grow as \cite{Arg02,BG}
\begin{equation}
\label{I_LZ_av}
\left< I_{LZ} \right> \sim 
\begin{cases}  \frac{h_{KS}}{\ln 2} \; t, & z < 2 \cr
h_{\alpha} \; t^{\alpha}, & z > 2,
\end{cases}
\end{equation}  
where $\alpha=1/(z-1)$. Numerical simulations shown in Fig.\ \ref{Manneville_LZ} are in agreement with Eq.\ (\ref{I_LZ_av}). The coefficient of proportionality is given by the Kolmogorov-Sinai entropy for $z < 2$ and by Krengel's entropy Eq.\ (\ref{h_alf}) calculated numerically (see the text bellow).
\begin{figure}[t]
\centerline{\psfig{figure=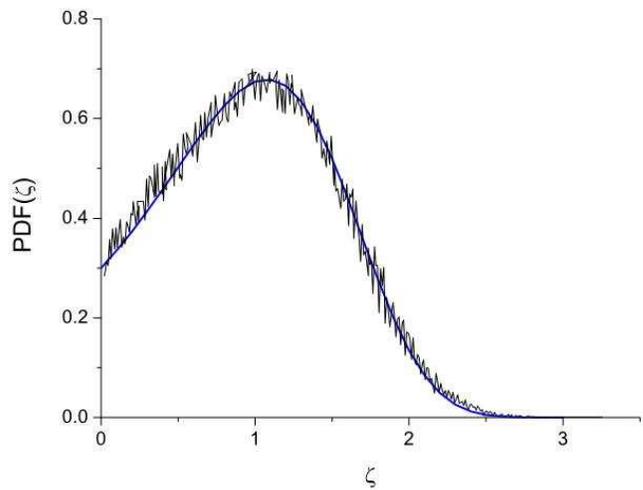,width=90mm,height=70mm}}
\caption{(color online) The PDF of $\zeta=C_{LZ}/\alpha \left< \lambda_{\alpha} \right>$ for the map in Eq.\ (\ref{map_eq_1}) with $\alpha = 0.75$ and $a = 1$. Here $10^5$ trajectories were iterated for $t = 10^5$. Smooth solid line corresponds to analytical density Eq.\ (26).}
\label{Manneville_LZ_PDF}
\end{figure}

Now, we define the complexity of the Lempel-Ziv algorithm as
\begin{equation}
\label{C_LZ_alf}
C_{LZ} = \lim_{t \rightarrow \infty} 
\begin{cases}  \frac{I_{LZ}}{t}, & z < 2 \cr
\frac{I_{LZ}}{t^{\alpha}}, & z > 2.
\end{cases}
\end{equation} 
In the ergodic phase, $z < 2$, $C_{LZ}$ converges to the limit and the average of complexity is equal to the Kolmogorov-Sinai entropy, see Eq.\ (\ref{C_LZ_ergod}). In the non ergodic phase, $z > 2$, $C_{LZ}$ does not converge to a constant but remains a random variable. For the map in Eq.\ (\ref{map_eq_1}) the distribution of the scaled variable $\zeta = C_{LZ}/\alpha \left< \lambda_{\alpha} \right>$ is given by the one-sided L{\'e}vy PDF Eq.\ (\ref{renewal_pdf_zeta}) as it was shown in Ref.\ \cite{SA06}. Numerically calculated PDF of $\zeta$ is shown in Fig.\ (\ref{Manneville_LZ_PDF}). The average complexity in this case is given by $\left< C_{LZ} \right> = h_{\alpha}$, where $h_{\alpha}$ is the Krengel's entropy \cite{Zwe06}. Using this the Krengel's entropy can be interpreted as an average speed of the information creation by the system. Combining it with Eq.\ (\ref{GPesin}), we get
\begin{equation}
\label{GPesin_LZ}
\left< C_{LZ} \right> = h_{\alpha} = \alpha \left< \lambda_{\alpha} \right>.
\end{equation} 
Generalized Pesin's identity Eq.\ (\ref{GPesin_LZ}) shows that the average speed of the information creation by the system is equal to the average speed of separation of trajectories for intermittent systems.

\section*{SUMMARY AND DISCUSSION}\label{Discus}

1) For the Pomeau-Manneville map Eq.\ (\ref{map_eq_1}) with one unstable fixed point \cite{GW88} and for the map with two unstable fixed points Eq. (2) the separation of trajectories is given by $dx_t \sim dx_0 e^{\lambda_{\alpha}t^{\alpha}}$. The PDF of scaled $\lambda_{\alpha}/\left< \lambda_{\alpha}\right>$ is given by one-sided L{\'e}vy PDF Eq.\ (\ref{renewal_pdf_zeta}) in agreement with ADK theorem \cite{JKP}.

2) For $0<\alpha<1$ ($z>2$) the measure is non normalizable and has infinite invariant density which is calculated semi analytically for the Pomeau-Manneville map Eq.\ (17) and numerically for the map with two unstable fixed points.

3) The average of $\lambda_{\alpha}$ Eq.\ (34) is found using the infinite invariant density for $0<\alpha<1$ ($z>2$). In contrast when $\alpha=1$ ($z<2$) the Lyapunov exponent is determined by the normalizable invariant density Eq.\ (4).

4) Standard concept of KS-entropy breaks down. The proper entropy is the Krengel's entropy of the first return map Eq.\ (36) \cite{Krengel}.

5) The scaled Lempel-Ziv complexity $C_{LZ}/\alpha \left< \lambda_{\alpha}\right>$ is a random variable distributed according to the one-sided L{\'e}vy law \cite{SA06}. The average complexity of the Lempel-Ziv algorithm is equal to the Krengel's entropy $\left< C_{LZ} \right> = h_{\alpha}$.

6) The Krengel's entropy and the average complexity of the Lempel-Ziv algorithm are related to the average $\left< \lambda_{\alpha}\right>$ Eq.\ (45).

7) Systems with zero Lyapunov exponents exhibit non-exponential separation of nearby trajectories \cite{Zasl,ZE05,Latora,DC01,ACG97,ARV04,Li,CP99,CNFV03,Gras}. Most of these examples have sub-exponential (namely, a power law) separation of trajectories, except the generalized logistic map close to tangent bifurcations, where trajectories separate faster than exponentially \cite{BR02}. It would be interesting to investigate possible generalized Pesin's identity for such systems.

(8) Very recently the distribution of time averaged mean square displacements of the stochastic continuous time
random walk (CTRW) processes was shown to be described by an equation similar to Eq.\ (27) \cite{He}. Relation of the CTRW model and intermittent dynamical systems is well established \cite{Zum93,BB06}. We expect that the distribution of time averaged mean square displacements in open intermittent systems \cite{GT84,Zum93} will follow Eq.\ (27). Thus, the universal fluctuations of generalized Lyapunov exponents (Fig. 6) and complexity of the Lempel-Ziv algorithm (Fig.\ 10) could be found also in open systems and for other observables. More work in this direction is needed.
 
\section*{ACKNOWLEDGMENTS}

This work was supported by the Israel Science Foundation. We thank J.~Aaronson, J.~Klafter, R.~Zweim\"uller, A.~Robledo and R.~Klages for discussions. NK thanks R.~Klages for having initiated this research as a supervisor during his Ph.D. thesis work by posing the problem to him. After this work was completed we were informed about the work of Klages and Howard which pursues a different approach to the problem of generalizing of Pesin's identity for intermittent systems \cite{KH}.

\subsection*{APPENDIX}

In this Appendix we solve the non-homogeneous Eq.\ (\ref{g_Man}). The Pomeau-Manneville map Eq.\ (\ref{map_eq_1}) has the form
\begin{equation}
x_{n+1} = M(x_n) = x_n + a x_n^z | \; mod \; 1, 
\end{equation}
with $a>0$ and $z>1$. The map is discontinuous at $x=\xi$, where $\xi + a \xi^z = 1$. The continuous approximation of this map valid for small $x_n$ is
\begin{equation}
\label{conteq}
\frac{dx}{dt} = a x^z, \; \; x \ll 1.
\end{equation}
Consider the following stochastic model of the Pomeau-Manneville map: a particle is moving in the laminar region $x \in (0, \xi)$ according to the equation of the motion Eq.\ (\ref{conteq}). At the boundary $x=\xi$ it is reinjected back to a random position in $(0, \xi)$. One can write the equation for the evolution for the probability density $\rho(x,t)$
\begin{equation}
\label{g_Man2}
\frac{\partial \rho(x,t)}{\partial t} = - \frac{\partial}{\partial x} (a x^z \rho(x,t)) + S(t),
\end{equation}
where $S(t)$ is chosen to fulfill conservation of normalization, namely $\int_0^{\xi} dx \; \rho(x,t)=const.$ independent of time
\begin{equation}
\int_0^{\xi} dx \; \frac{\partial \rho(x,t)}{\partial t} = - \int_0^{\xi} dx \; \frac{\partial}{\partial x} (a x^z \rho(x,t)) + S(t) \int_0^{\xi} dx,
\end{equation}
which gives 
\begin{equation}
\label{St}
S(t) = a \xi^{z} \rho(\xi,t).
\end{equation}
$S(t)$ is a source term which according to Eq.\ (\ref{St}) depends on the probability of finding a particle at time $t$ on $x=\xi$.

Now we solve the non-homogeneous Eq.\ (\ref{g_Man2}) with $0<x<\xi$ and with initial condition $\rho(x,0)=\xi^{-1}$. It is easy to verify that the solution of the corresponding homogeneous equation with $0<x<\xi$ is
\begin{equation}
\label{rhoh}
\rho_h(x,t) = \frac{\xi^{-1}}{\left[ 1 + a(z-1)tx^{z-1} \right]^{\frac{z}{z-1}}}.
\end{equation}
Hint: Eq.\ (\ref{conteq}) has the solution 
\[ x(t) = \left[ x_{0}^{1-z} - a(z-1)t \right]^{\frac{1}{1-z}}, \] 
which is valid for not too long time $t<x_0^{1-z}/(a(z-1))$. Assuming the uniform initial distribution of $x_0$, one find the density of particles for $x\ll1$ in the form of Eq.\ (\ref{rhoh}). This density is not normalized since it does not include the effect of the source $S(t)$.

Let us rewrite Eq.\ (\ref{rhoh}) in the form 
\begin{equation}
\rho_h(x,t) = \xi^{-1} O_x(t),
\end{equation}
where 
\begin{equation}
O_x(t) = \frac{1}{\left[ 1 + a(z-1)tx^{z-1} \right]^{\frac{z}{z-1}}}.
\end{equation}
The Laplace transform of $O_x(t)$ is given by Eq.\ (\ref{O_s}). According to Duhamel's principle, a particular solution of the non-homogeneous equation is given by the convolution of the homogeneous equation with function $S(t)=a \xi^{z} \rho(\xi,t)$ 
\begin{equation}
\rho_p(x,t) = \int_0^t d \tau S(\tau) \rho_h(x,t-\tau).
\end{equation}
The general solution of the non-homogeneous equation Eq.\ (\ref{g_Man2}) is the sum of the particular solution of the non homogeneous equation and the solution of homogeneous equation
\begin{equation}
\rho(x,t) = \rho_p(x,t)+\rho_h(x,t).
\end{equation}
Taking the Laplace transform of the solution, we get
\[\tilde{\rho}(x,s) = \tilde{\rho}_p(x,s) + \tilde{\rho}_h(x,s) = \]
\begin{equation}
\label{nonhsol}
= a\xi^{z-1}\tilde{\rho}(\xi,s)\tilde{O}_x(s)+\xi^{-1}\tilde{O}_x(s).
\end{equation}
For $x=\xi$ one find
\begin{equation}
\tilde{\rho}(\xi,s) = \frac{\xi^{-1}\tilde{O}_{\xi}(s)}{1-a \xi^{z-1}\tilde{O}_{\xi}(s)}.
\end{equation}
Substituting this into Eq.\ (\ref{nonhsol}), we finally obtain solution presented in Eq.\ (\ref{g})
\begin{equation}
\label{Eq61}
\tilde{\rho}(x,s) = \frac{\xi^{-1}\tilde{O}_{x}(s)}{1-a \xi^{z-1}\tilde{O}_{\xi}(s)}.
\end{equation}
Using Eq.\ (\ref{O_s}), we find
\begin{equation}
\int_0^{\xi} dx \tilde{O}_x(s) = \frac{1}{s} \left( \xi - a \xi^z \tilde{O}_{\xi}(s) \right),
\end{equation}
which, together with Eq.\ (\ref{Eq61}), allows to check for the normalization
\begin{equation}
\int_0^{\xi} dx \tilde{\rho}(x,s) = \frac{1}{s}.
\end{equation}



\begin{thebibliography}{10}
\bibitem{note}
In the literature two terms weak chaos and pseudochaos are used to describe random systems with zero Lyapunov exponent \cite{Zasl,Kla}.
\end{thebibliography}
\end{document}